\documentclass{elsart}
\newcommand{\be}{\begin{eqnarray}}
\newcommand{\en}{\end{eqnarray}}
\begin{document}
\begin{frontmatter}
\title{A Field Theoretic Investigation of Spin in QCD}
\author{Asmita Mukherjee}
\address{Saha Institute of Nuclear Physics\\1/AF, Bidhannagar, Calcutta
700064, India}
\date{3rd August, 2000}
\begin{abstract}
Utilizing the kinematical boost in light-front formalism one
can address the issue of relativistic spin operators in an arbitrary
reference frame. In the gauge $A^+=0$, the interaction dependent transverse
spin operators can be separated into three parts. In analogy with the
helicity sum rule, we propose a transverse spin sum rule. We perform a one
loop renormalization of the transverse spin operator and show that the
counterterm needed is the same as the linear mass counterterm in the
light-front QCD Hamiltonian.  
\end{abstract}
\end{frontmatter}

\section{Introduction}
The issue of the relativistic spin operators in quantum field theory in an
arbitrary reference frame is quite complex. The spin operators can be
constructed from the 
Pauli-Lubanski operators, however, they are interaction dependent in an
arbitrary frame and it is 
quite difficult to separate the center of mass motion from internal motion 
for a composite system in the usual equal time formulation \cite{bak}.
Light-front formulation of quantum field theory is more suitable
for the study of relativistic spin operators since boost is kinematical in
this formulation. The interaction dependence of the light-front transverse spin
operators come from the transverse rotation generators, which are dynamical
(interaction dependent) in light-front theories in contrast to the
equal-time case. The separation of the center of mass motion from
the internal motion in this case is as simple as in non-relativistic theory.

\section{Light-Front Spin Operators}
In terms of Poincare generators, light-front spin operators can be written
as \cite{ls78},
\be
M T^i &&= W^i - P^i T^3,~~~~~~~~~~~~i=1,2
\nonumber\\
&&= \epsilon^{ij} ( {1\over 2}F^jP^+ -{1\over 2}E^jP^- + K^3P^j)-P^iT^3,
\en
where $W^\mu$ is the Pauli-Lubanski operator, $W^\mu = -{1\over 2}
\epsilon^{\mu \nu\rho \sigma} M_{\nu \rho} P_\sigma$ with $\epsilon^{+-12} =
-2$, $M$ is the mass operator;
$M^{\mu \nu}$ are the generalized angular momentum and $P^\mu$ is the
momentum 4-vector. The boost operators are $M^{+-} = 2K^3, M^{+i} = E^i$.
Rotation operators are, $M^{12} = J^3, M^{-i} = F^i$. The third component of
the spin operator is the helicity operator which 
is defined as,
\be
T^3 = {W^+\over P^+} = J^3 + {1\over P^+}( E^1 P^2 - E^2 P^1).
\en
In addition to the light-front Hamiltonian $P^-$, the transvese rotation
operators $F^i$ are dynamical (interaction dependent).
As a result, the transverse spin operators are also interaction dependent.
 
In the light-front gauge $A^+ = 0$, the helicity operator is equal to its
naive canonical form, independent of interaction, provided the fields vanish
at infinity. This situation can be contrasted with the third component of
the spin operator in the equal-time case, which is interaction dependent.
 Explicitly, the light-front helicity operator can be separated into quark
and gluon orbital and intrinsic parts at the operator level and the helicity
sum rule for the nucleon is given by \cite{hk,brod},
\be
\langle P S^{||} \mid T^3_{f(i)} + T^3_{f(o)} +
T^3_{g(o)} + T^3_{g(i)}  \mid P S^{||} \rangle 
= \pm {1\over 2}.
\en
It can be verified that the matrix element of $ T^3_{f(i)}$ is
directly related to the first moment of the flavor singlet part of the
longitudinally polarized structure function $g_1(x, Q^2)$.

In order to construct the transverse spin operators, one has to calculate
the dynamical transverse rotation operators. The transverse spin operators
 cannot be separated into orbital and spin
parts. We have shown that \cite{let,pap1}, in the light-front gauge, there exists a physically interesting
decomposition for them, 
\be
T^i = T^i_{I} + T^i_{II} + T^i_{III},
\en
where $T^i_{II}$ and $T^i_{III}$ do not depend on the
coordinates explicitly and arise from the fermionic and gluonic parts of the
gauge invariant, symmetric energy momentum tensor respectively. In analogy
with the helicity sum rule, we propose a transverse spin sum rule for the
nucleon,
\be
\langle P S^\perp \mid T^i_{I} + T^i_{II} + T^i_{III} \mid P S^\perp \rangle
= {1\over 2}.
\label{sum}
\en
The intrinsic fermionic part $T^i_{II}$ is directly related to the first
moment of the flavor singlet part of the structure function $g_T$ measured 
in transverse polarized
scattering. The difference from the helicity case is the non-trivial
interaction dependence of the transverse spin operators, because of which the
above relation has no partonic interpretation. 

\section{Renormalization of the Transverse Spin Operators}
The transverse spin operators acquire divergences in perturbation theory and
have to be renormalized. We have performed one loop renormalization of the
full transverse spin operator by evaluating its matrix element for a quark
state dressed with a gluon in light-front Hamiltonian perturbation theory
\cite{pap2}.
Explicit evaluation in an arbitrary reference frame shows that all the center of mass momentum dependence
get canceled. Also the contribution from $T^i_I$ exactly cancels the
contribution from $T^i_{III}$ and the entire contribution comes from
$T^i_{II}$, which is given by,
\be
\langle P, S^i \mid M T^i \mid P, S^i \rangle =
{m\over 2} \left ( 1 + {3 \alpha_s \over {4\pi}} C_f ln{Q^2\over \mu^2}
\right),
\en
where $m$ is the bare mass of the quark, $C_f$ is the color factor and $\mu$
is the hadronic factorization scale. The details of the calculation can be
found in \cite{pap2}. Using the expression of the
renormalized mass of the quark in terms of the bare mass \cite{hari3},    
\be
m_R = m \left( 1 + { 3 \alpha_s \over {4 \pi}} ln {Q^2 \over \mu^2} \right
),
\en
we verify the sum rule (\ref{sum}). This shows that only one counterterm 
is needed to renormalize the transverse spin operator, and that is the same
as the linear mass counterterm in the light-front QCD Hamiltonian. Also,
the mass of the quark plays a very important role in this case since the
transverse spin operator is responsible for helicity flip and it is the
terms linear in mass that causes helicity flip in light-front theory. 

In summary, we have studied the relativistic spin operators for a composite system in
an arbitrary reference frame in light-front QCD. The transverse spin
operators are interaction dependent and in the light-front gauge, they can
be separated into three physically interesting parts, provided the fields
vanish at the boundary. The helicity operator in the same gauge is
interaction independent. In analogy with the helicity sum rule, we have
proposed a transverse spin sum rule. We have renormalized the transverse
spin operators upto one loop and shown that the counterterm that has to be
added is the same as the linear mass counterterm in light-front QCD
Hamiltonian.

\section{Acknowledgments}

This is based on the work done in collaboration with A. Harindranath  and R.
Ratabole. I would also like to thank the organisers of X-th International
Light-Cone Meeting on Non-Perturbative QCD and Hadron Phenomenology for
giving me the chance to present my work.

\end{document}